\documentclass[amsmath,amssymb]{revtex4-1}
\usepackage{graphicx}
\usepackage{mathptmx}      

\usepackage{dcolumn}   
\usepackage{bm}        
\usepackage{amssymb}   
\usepackage{epsfig}
\usepackage{float}
\usepackage{epsf}
\usepackage{colordvi}

\usepackage{newtxtext,newtxmath}

\newcommand{\reference}[5]
{\sl #1\ \rm (#4) #2 #3:#5}

\begin{document}

\title{Range-separated density-functional theory applied to the beryllium dimer and trimer}

\author{Peter Reinhardt}\email{Peter.Reinhardt@upmc.fr}
\author{Julien Toulouse}
\author{Andreas Savin}

\affiliation{Laboratoire de Chimie Th\'eorique (LCT), Sorbonne Universit\'e and CNRS, 4 place Jussieu, F-75252 Paris, France}

\date{September 17, 2018}

\begin{abstract}
The beryllium dimer and trimer are, despite their small number of electrons, excellent systems for assessing electronic-structure computational methods.
With reference data provided by multi-reference averaged coupled-pair functional calculations, we assess several variants of range-separated density-functional theory, 
combining long-range second-order perturbation theory or coupled-cluster theory with a short-range density functional. The results show that 
i) long-range second-order perturbation theory is not sufficient, ii) long-range coupled-cluster theory gives reasonably accurate potential energy curves, but iii) provided 
a relatively large value of $\mu=1$ bohr$^{-1}$ for the range-separation 
parameter is used. The article is dedicated to the memory of J\'anos G. 
\'Angy\'an.
\end{abstract}

\maketitle

\section{Introduction}
\label{intro}

Density-functional theory (DFT) has become in the last 20 years a widely used
tool in quantum chemistry due to its good performance for structural and thermochemical
properties at an advantageous cost even for medium-size molecules and
larger entities. Nevertheless, research is still very active to improve the 
usual DFT approximations, notably for the treatment of long-range dispersion 
interactions and the treatment of genuine multi-reference cases. 

To address the former issue, J\'anos \'Angy\'an, to the memory of whom this 
article is dedicated, worked on the development of range-separated 
density-functional theory (RS-DFT), in a long-standing and 
fruitful collaboration with the present authors (see, e.g., Refs.~\cite{rangeseparate,ToulouseRPA,ToulouseCCD,Chermak,Mussard}). 
A never-published work that was started with J\'anos some years ago was on the 
application of RS-DFT on the dimer and trimer of Beryllium, which contain both 
dispersion and multi-reference effects. For this special issue, we have 
reexamined this work and completed it.

Be$_2$ is weakly bound, and has been the subject to a long-standing discussion 
between theoreticians and experimentalists  
\cite{JonesBe2,LiuMcLean,BeExp,Roeggen,PatkowskiBeMgCa,Ruedenberg,%
Patkowski,ExpBeII}, 
whereas Be$_3$ is a fairly stable aggregate, thus of completely 
different nature in bonding. This has been reviewed by e.g.\ Kalemos in a 
recent publication \cite{Kalemos}. The goal of the present paper is to see 
whether different flavors of RS-DFT can correctly 
describe these different chemical bonds.

The paper is organized as follows. After a short recall of RS-DFT, we give a 
brief survey of existing data, and construct consistent reference results for 
both the dimer and the trimer. To these we compare different flavors of RS-DFT 
and discuss their performance. All further technical details are collected in 
the Appendix, and the underlying data sets can be consulted in the 
Supplementary Material.   

\section{Range-separated density-functional theory}

The common Kohn-Sham procedure \cite{KohnSham} minimizes the following energy 
expression with a single Slater determinant wave function $\Phi$
\begin{equation} E_{\rm exact} = \min_{\Phi}\left\{ 
\langle\Phi|\hat{T}+\hat{V}_{\rm ne}|\Phi\rangle + E_{\rm Hxc}\left[
  n_{\Phi}\right]
\right\}, \end{equation}
where $\hat{T}$ is the kinetic energy operator, $\hat{V}_{\rm ne}$ is the 
electron-nucleus attraction operator,
and $E_{\rm Hxc}\left[n_{\Phi}\right]$ is the Hartree-exchange-correlation 
functional evaluated at the density $n_\Phi$ produced by $\Phi$. With the exact 
density functional $E_{\rm Hxc}[n]$, the exact ground-state energy would be 
obtained, in the limit of a complete basis set, by virtue of the Hohenberg-Kohn 
theorem.

The Kohn-Sham method can be considered as a special case of a more general 
RS-DFT scheme (see, e.g., Ref.~\cite{Colonna}) 
\begin{equation} E_{\rm exact} = \min_{\Psi}\left\{ 
\langle\Psi|\hat{T}+\hat{V}_{ \rm ne }+\hat{W}_{\rm ee}^{\rm lr}|\Psi\rangle +
E^{\rm sr }_{\rm Hxc }\left[n_{\Psi}\right]  
\right\}, \label{eq:sr} \end{equation}
where $\Psi$ is a general, multi-determinant wave function,
$\hat{W}_{\rm ee}^{\rm lr} = \sum_{i<j} w_{\text{ee}}^{\rm lr}(r_{ij})$ is a 
long-range electron-electron interaction and  
$E^{\rm sr}_{\rm Hxc}[n]$ the associated short-range complement
Hartree-exchange-correlation density functional. The range separation of the 
electron-electron interaction    
\begin{equation} \frac{1}{r_{ij}} = w_{\rm ee}^{\rm lr}(r_{ij}) +   \left( 
\frac{1}{r_{ij}}-  w_{\rm ee}^{\rm lr}(r_{ij}) \right),
\end{equation}
is achieved through the use of the error function with an arbitrary parameter
$\mu$ 
\begin{equation}  w_{\rm ee}^{\rm lr} (r_{ij}) \ = \ \frac{{\rm erf}(\mu\,
  r_{ij})}{r_{ij}}. \end{equation}


In practice, approximations must be used for the wave function $\Psi$.
A first step is to use only a single-determinant wave function $\Phi$, which
leads to the range-separated hybrid (RSH) approximation \cite{rangeseparate}
\begin{equation} E_{\rm RSH} = \min_{\Phi}\left\{ 
  \langle\Phi|\hat{T}+\hat{V}_{\rm ne} +
  \hat{W}_{\rm ee}^{\rm lr}|\Phi\rangle    + 
  E^{\rm sr}_{\rm Hxc}\left[n_{\Phi}\right]
\right\}, 
\label{eq:RSH}
\end{equation}
the expectation value of the long-range interaction, $\langle\Phi|\hat{W}_{\rm 
ee}^{\rm lr}|\Phi\rangle$, giving a long-range Hartree-Fock (HF) contribution 
to the energy. The minimization in Eq.\ (\ref{eq:RSH})\ leads to self-consistent
Kohn-Sham-like RSH equations for the orbitals $|\phi_i \rangle$ and orbital 
energies $\epsilon_i$ 
\begin{equation}
 \left( \hat{T} + \hat{V}_{\rm ne} + \hat{V}_{\rm H} +
  \hat{V}_{\rm x,HF}^{\rm lr} + 
  \hat{V}_{\rm xc}^{\rm sr} \right)\,|\phi_i\rangle\ = \
\epsilon_i\, |\phi_i\rangle,
\label{eq:RSH_II}
\end{equation}
where $\hat{V}_{\rm H}$ is the full-range Hartree potential, $\hat{V}_{\rm 
x,HF}^{\rm lr}$ is the long-range nonlocal HF exchange potential (evaluated 
with the ``erf'' part of the electron-electron interaction), and $\hat{V}_{\rm 
xc}^{\rm sr}$ is the the short-range exchange-correlation potential (obtained 
from the functional derivative of $E^{\rm sr}_{\rm xc}[n]$).

\begin{figure}
  \includegraphics[width=\textwidth]{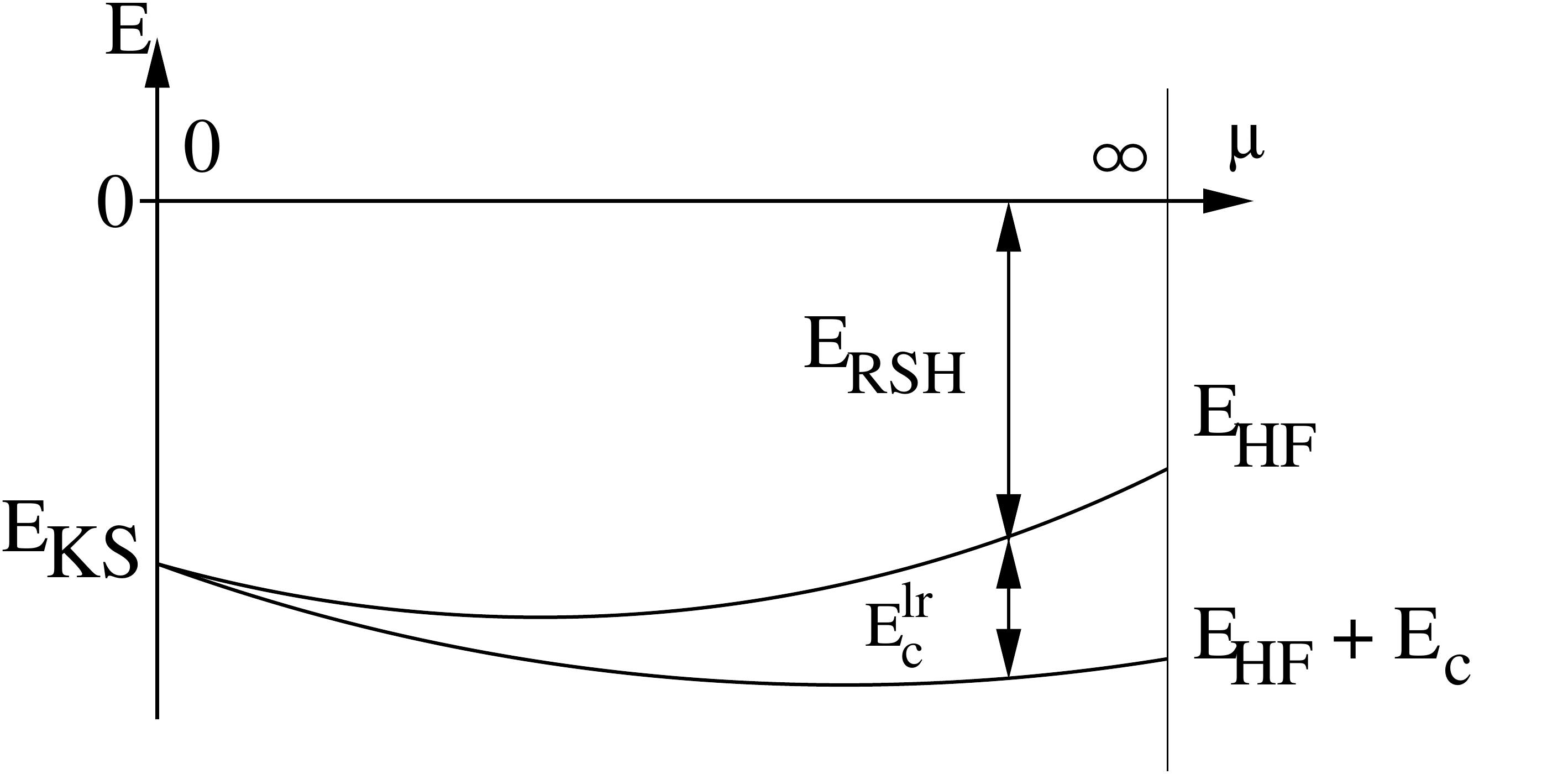}
\caption{Schematical composition of the RS-DFT total energy as the RSH
  energy plus the long-range correlation energy. For $\mu=0$, one obtains
  standard Kohn-Sham theory without explicit long-range correlation
  correction, and for $\mu\rightarrow\infty$ RSH reduces to HF (no
  short-range functional) to which the correlation energy is added.}
\label{fig:adiab}  
\end{figure}

Eq.\ (\ref{eq:RSH})\ does not yet include the
long-range correlation energy. The exact energy is formally the sum of the RSH 
energy and the long-range correlation energy: 
\begin{equation}
E_{\rm exact} = E_{\rm RSH} + E^{\rm lr}_{\rm c} 
\end{equation}
as depicted schematically in Fig. \ref{fig:adiab}.
Whereas for instance dispersion interactions are often added to DFT via 
{\sl ad hoc} corrections based on the evaluation of atomic
polarizabilities \cite{Grimme,Elstner,Johnson}, the RS-DFT formalism allows us
to include explicitly these important contributions to intermolecular
interactions, only based on (i) the single parameter of the range separation
$\mu$, (ii) the choice of the short-range exchange-correlation functional, and 
(iii) the long-range correlation method. The long-range correlation energy can 
be calculated by second-order M\o{}ller-Plesset (MP2) perturbation 
theory~\cite{rangeseparate}, which leads to exactly the same equations as 
standard MP2 \cite{MollerPlesset}, i.e., for closed-shell systems:
\begin{equation}
E^{\rm lr}_{\rm c,MP2} =
\sum_{ijab}\frac{[2(ia|jb)^{\rm lr}-(ib|ja)^{\rm lr}](ia|jb)^{\rm 
lr}}{\epsilon_i + \epsilon_j-\epsilon_a-\epsilon_b},    
\end{equation}
where $i,j$ and $a,b$ refer to occupied and virtual RSH orbitals, respectively, 
and $(ia|jb)^{\rm lr}$ are the long-range two-electron integrals (using the 
``erf'' interaction). Of course, all these quantities (integrals and orbital 
energies) depend on the range-separation parameter $\mu$. Similarly,
the long-range correlation energy can be calculated by coupled cluster singles 
doubles and perturbative triples (CCSD(T))~\cite{Goll,ToulouseCCD}. 

One of the central advantages of these methods with respect to standard
quantum chemistry approaches like MP2 or coupled-cluster theory is the
considerably weaker basis-set dependence as the short-range correlation hole is 
taken already into account via the density-functional part. As a consequence, 
basis-set superposition errors (BSSE) should be significantly reduced with 
respect to standard wave-function-based correlation calculations.

Concerning the short-range exchange-correlation functional, several
approximations \cite{Colonna,Goll,Goll2,SRfunctionals,ColonnaII,Paziani,muhalfFro,GollTPSS}
have been proposed. We employed from the different possibilities the
short-range local-density approximation (LDA) of Ref. \cite{Paziani} and the 
short-range Perdew-Burke-Ernzerhof (PBE) functional of Ref. \cite{Goll2}.  We 
will refer to the range-separated methods as RSHLDA+MP2 and RSHLDA+CCSD(T),
and RSHPBE+MP2 and RSHPBE+CCSD(T). Of course, setting the range-separation 
parameter $\mu$ to infinity, we obtain the standard MP2 or CCSD(T) 
approach without a density-functional part.

\section{Brief review of existing data}
In the past, Be clusters have been abundantly studied, and an 
outstanding object of interest is the dimer (for a recent review see 
\cite{BeReview}). Compared to the heavier 
alkaline-earth dimers, Be$_2$ already shows a shorter and stronger binding than 
expected for a purely dispersive interaction with an interatomic equilibrium 
distance and cohesion energy of about 2.4$\,$\AA\ and 2.5$\,$kcal/mol, 
respectively. 

This binding energy and interatomic distance has been a subject of long
debates, experimentally and theoretically (see Refs.\
\cite{JonesBe2,LiuMcLean,BeExp,Roeggen,PatkowskiBeMgCa,Ruedenberg,Patkowski}).
The experimental difficulties are summarized in a recent article~\cite{ExpBeII},
and from the theoretical side both the near degeneracy of the
occupied 2$s$ orbital with the unoccupied $2p$ orbitals~\cite{HandyI,Lepetit}
and the shallowness of the potential well without being a purely dispersion
interaction require a careful treatment of electron correlation 
\cite{BeReview}. 

It is therefore not surprising that at the HF level, Be$_2$ is not
bound at all. Single-reference  
configuration-interaction (CI) schemes, truncated at the level of single and
double excitations (CISD) are not suited for this system \cite{HandyI,Zhang},
even when including all exclusion-principle-violating (EPV) diagrams in a
coupled-electron-pair approximation (called full CEPA or self-consistent
size-consistent CI) \cite{SCSC}. One has to go indeed to multi-reference or
coupled-cluster (including triple excitations \cite{Bartlett,Zhang})
correlation methods for a correct description. 

With the work of R\o{}eggen and Veseth \cite{Roeggen}, Patkowski \textit{et al.}\
\cite{Patkowski}, Schmidt \textit{et al.}\ \cite{Ruedenberg}, and corresponding 
experimental data \cite{ExpBeII}, we can consider that the potential well of 
the Be dimer is precisely known. DFT with usual semilocal approximations 
overestimates the interaction energy, however the work of Jones \cite{JonesBe2} 
was the first theoretical one to claim the Be dimer to be significantly bound.  
RS-DFT with long-range MP2 or random-phase-approximation (RPA) approaches was 
applied to the Be dimer~\cite{BeAngyan,ToulouseRPA,ToulousePRL}, giving 
significantly underestimated interaction energies.
Range-separation in combination with multi-reference perturbation theory 
(NEVPT2) has been published by Fromager \textit{et al.} \cite{Fromager} with the 
conclusion that differences to the reference data despite the 
multi-reference long-range correlation treatment should be imputed to 
deficiencies of the employed short-range PBE functional. 

For Be$_3$ in D$_{3h}$ symmetry HF theory gives
no overall binding \cite{HandyII,Taylor,Kaplan,Klopper,Stefano}. Nevertheless, a
distinct local minimum is produced, contrary to Be$_2$ where the HF
potential curve is purely repulsive. Including electron
correlation \cite{HandyII,Taylor,Kaplan,Klopper,Stefano} results in a cohesion
energy of the triatomic ``molecule'' between 15 and 30$\,$kcal/mol, with an
equilibrium interatomic distance of about 2.2$\,$\AA, which is about the same
as for the local minimum in HF. The significant difference in the
binding of the dimer and the trimer (i.e.\ the importance of
non-additivity or ``3-body interactions'') has been discussed by Novaro and
Ko\l{}os \cite{Novaro} and Daudey {\sl et al.} \cite{DaudeyBe} already in the
seventies in the framework of HF theory. 
A comparative study \cite{Jae} of the dimer, trimer and tetramer of Be and of
Mg, based on MP2, CCSD(T), and DFT methods estimates the utility of different basis-set
extrapolation formulae and concludes that an extrapolation from the difference
between double-zeta and triple-zeta basis sets is sufficient to estimate
converged results for binding energies. As in the case of rare-gas dimers and 
the Be dimer, the LDA overestimates the binding energy of the trimer 
\cite{Khanna,LDABe3}. Even gradient-corrected functionals such as BPW91
result in too high binding energies \cite{Jellinek}, the effect being less 
important but still present for the B3LYP hybrid functional \cite{Beyer}.

\section{Results and discussion}
In order to dispose of a coherent reference data set for the two systems, we 
perform MP2, MP4, CCSD(T), MRCI, MR-ACPF \cite{Gdanitz}, and MR-AQCC \cite{AQCC} 
calculations with the MOLPRO program package \cite{Molpro}. For the 
multi-reference calculations two electrons and four orbitals of each Be atom 
are included in the complete active space (CAS), forming the reference space to 
which single and double excitations in the size-consistency-corrected ACPF or 
AQCC correlation formalism are added.

We employ the aug-cc-pV$X$Z basis sets ($X$ = D, T, Q --- see appendix), and for the
reference calculations we extrapolate the correlation energy via the inverse-cubic scheme 
$E_{\text{c}}(X)=E_{\text{c}}(\infty)+B/X^3$ to 
an estimate of the complete (valence) basis set limit. For 
RS-DFT calculations, however, it has been shown that 
an exponential form $E_{\text{c}}(X)=E_{\text{c}}(\infty)+B\exp(-\beta X)$ is 
more adequate \cite{Odile} -- we will therefore employ this scheme for the RS-DFT 
results. All data are corrected for the BSSE by the counterpoise scheme
of Boys and Bernardi \cite{BoysBernardi}, even though this correction remains 
always smaller than 0.2$\,$kcal/mol. The HF energy is
considered to be converged with the quadruple-zeta basis set.

For the wave-function-based correlation calculations, the core orbitals were
kept frozen, i.e.\ only the valence shells are explicitly correlated. Indeed, 
this choice is consistent with the fact that 
the aug-cc-pV$X$Z basis sets do not contain functions to correlate
core electrons. In the short-range 
functional, however, core correlation is automatically included even when long-range correlation is 
treated by a frozen-core wave-function-based method. 

In order to study core-valence correlation explicitly, we did some 
exploratory calculations with the core-polarization potentials (CPPs) of the 
Stuttgart group \cite{CPP} added to the frozen-core calculations, as well as 
correlated all-electron calculations in an aug-cc-pwCVTZ basis set with 
core functions \cite{Peterson} added to the standard aug-cc-pVTZ set.

\subsection{Wave-function-based reference calculations}


For the dimer the extrapolated MR-ACPF and MR-AQCC interaction energy curves
are close to the reference potential of Ref.\ \cite{Roeggen} or Ref.\ 
\cite{Patkowski} 
(see Table \ref{tab:be2-be3-WF}, left column for bond lengths and potential 
depths), while the CCSD(T) curve has a similar shape but underestimates the 
interaction energy significantly (see Supplementary material for data and 
graph). As discussed above, CCSD is not at all adequate here, and perturbation 
theory yields as well quite different interaction potentials, 
MP2 underestimating at the equilibrium distance and MP4 overestimating at 
larger distances.

\begin{figure*}[h!]
  \includegraphics[width=0.90\textwidth]{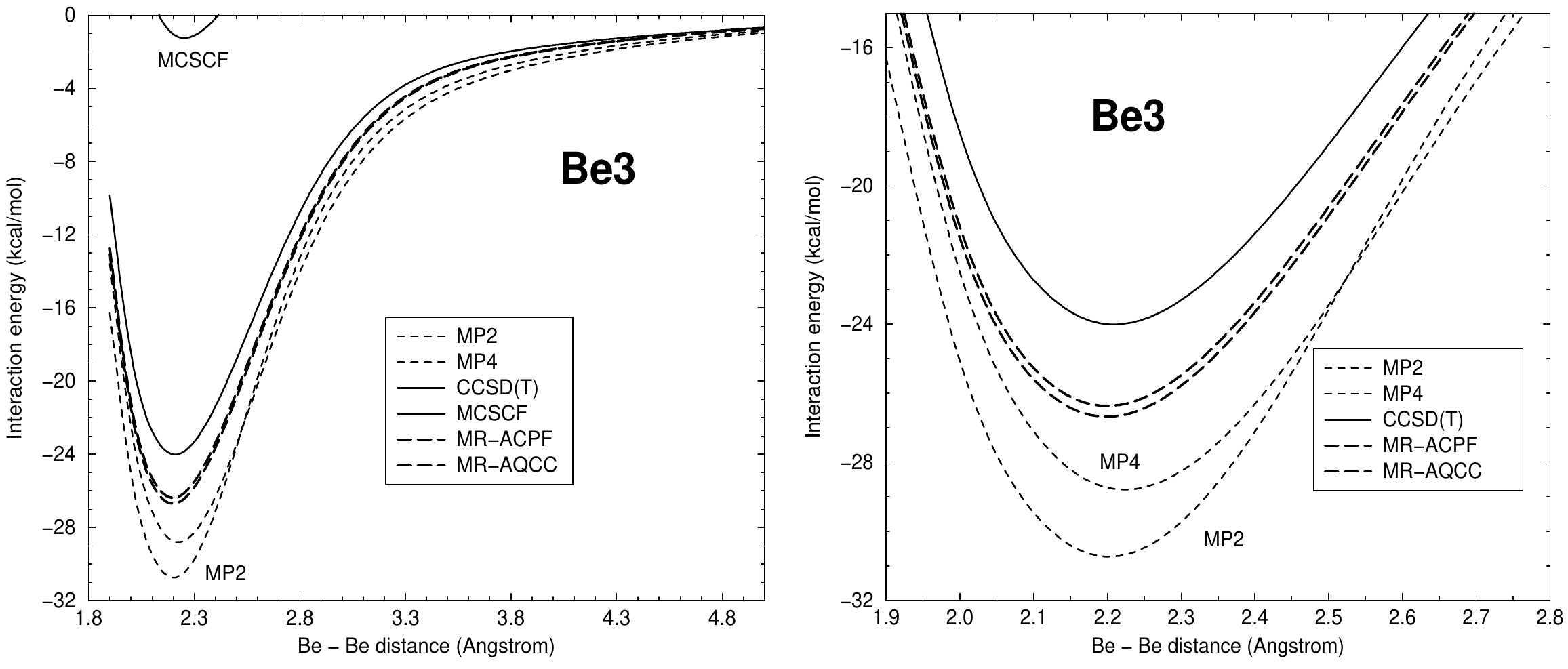}
\caption{Basis-set extrapolated interaction energy curves for
  the Be trimer, as obtained by wave-function-based methods. The right panel is 
a zoom into the minimum-energy region of the left panel.}
\label{fig:Pot_Be3_bw}  
\end{figure*}

For the trimer (see Figure \ref{fig:Pot_Be3_bw}) we do not dispose of an
accurate reference curve from the literature, but we see that CCSD(T) and
multi-reference methods give potentials of similar shape, and again
second-order perturbation theory yields significantly different results, but in 
the opposite way with respect to Be$_2$. MP4 overestimates the interaction 
energy as much as CCSD(T) underestimates it (Table \ref{tab:be2-be3-WF}, right 
column). CCSD again largely underbinds.

\begin{table}[h]
 \centering
  \caption{Equilibrium bond lengths (\AA) and interaction energies (kcal/mol) 
    from inverse-cubic basis-set extrapolated potential curves
    for the Be dimer and trimer. See text for the extrapolation procedure.}
  \begin{tabular}{l|rr|rr}
    \hline\hline
    &   \multicolumn{2}{c|}{Be$_2$}  & \multicolumn{2}{c}{Be$_3$} \\
    & $r_{\rm Be-Be} $ & $E_{\rm min}$   & $r_{\rm Be-Be}$ & $E_{\rm min}$ \\
    \hline
    MP2     & 2.71 & $-$1.32              & 2.20 & $-$30.7 \\
    MP4     & 2.52 & $-$2.38              & 2.22 & $-$28.8 \\
    CCSD    & 4.42 & $-$0.17              & 2.21 & $-$14.4 \\
    CCSD(T) & 2.48 & $-$1.87              & 2.21 & $-$24.0 \\ 
    MRCI    & 2.47 & $-$2.40              & 2.20 & $-$25.9 \\  
    MR-ACPF & 2.46 & $-$2.52              & 2.20 & $-$26.7 \\  
    MR-AQCC & 2.46 & $-$2.44              & 2.20 & $-$26.4 \\ 
    \hline\hline
  \end{tabular}
  \label{tab:be2-be3-WF}
\end{table}

By taking into account core and core-valence correlations, the interaction
energy is lowered (in a triple-zeta basis set) by about 0.1$\,$kcal/mol for the 
all-electron calculations, and by about 0.4$\,$kcal/mol using the CPPs, for the 
dimer. For the Be trimer the effect is about 1$\,$kcal/mol for the inclusion 
of core correlations, and about 4$\,$kcal/mol for the use of the CPPs. We thus 
use the frozen-core data, and, due to these estimations, we keep in mind that 
the true interaction potential may be slightly lower than our actual 
multi-reference data.

\subsection{RS-DFT calculations}
\begin{figure}[h!]
  \includegraphics[width=\textwidth]{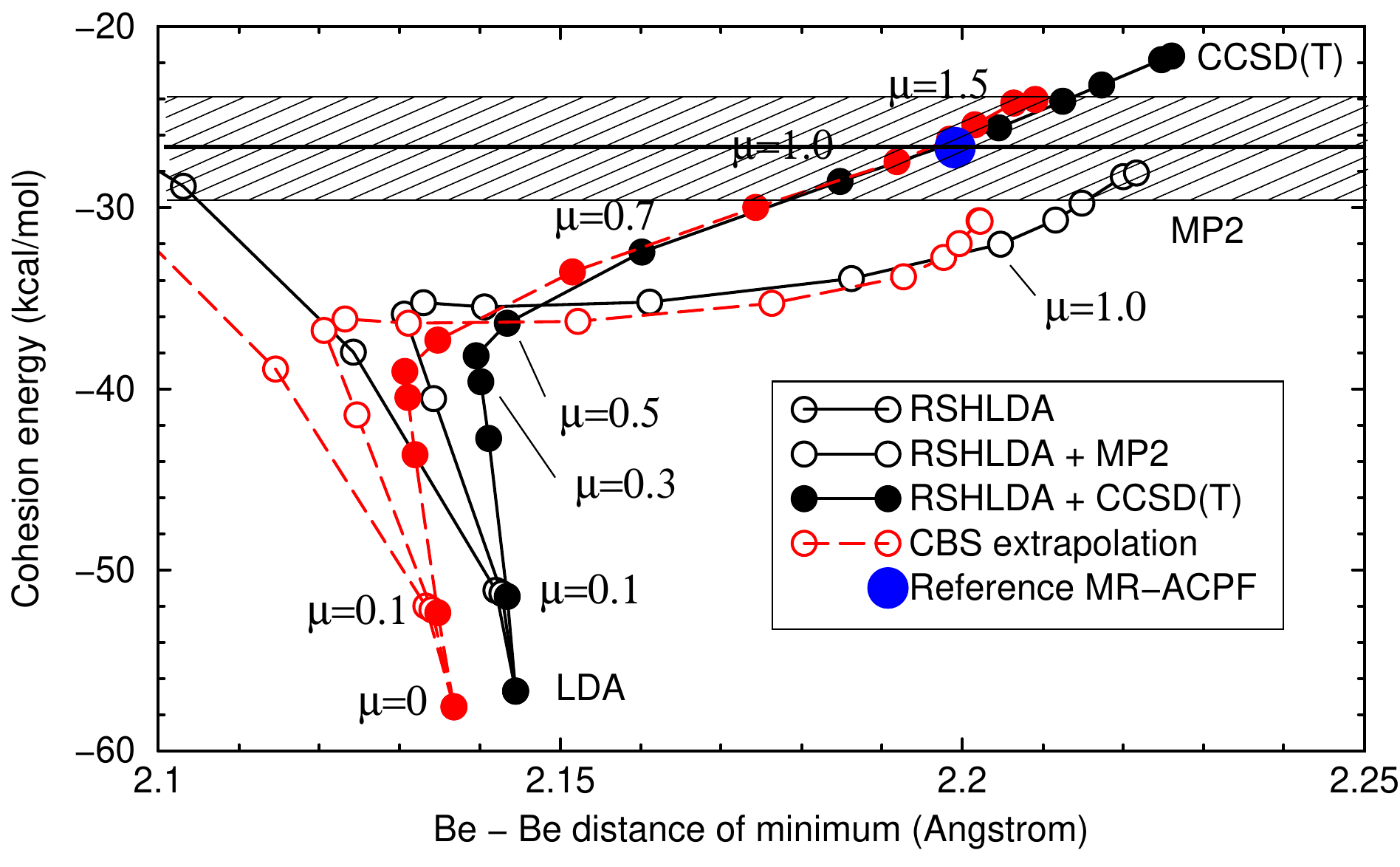}
\caption{Effect of the basis-set extrapolation for the equilibrium 
cohesion energies and bond lengths as a function of the 
range-separation parameter $\mu$, for Be$_3$. Full lines are for the 
aug-cc-pVTZ basis set, and dashed lines represent the corresponding 
extrapolated values.}
\label{fig:RSH-mu_bw2}  
\end{figure}

From the previous section we conclude that today's ``gold standard'' CCSD(T) 
with basis-set extrapolation gives for the trimer too low a cohesion energy, 
by about 2.5 kcal/mol or 10\%. The calculations need a non-negligible 
effort, which may be avoided by resorting to DFT-based calculations while 
resulting in not worse an error. In Figure \ref{fig:RSH-mu_bw2} we display the 
evolution of the RS-DFT interaction energies with the range-separation 
parameter $\mu$, in comparison to our MR-ACPF reference data, and the ``region 
of confidence'' around the reference energy as shaded area, which we chose 
to be the deviation of basis-set extrapolated CCSD(T) around the MR-ACPF 
reference. In order to remain inside this region, we may take a value for the 
range-separation parameter $\mu$ between 0.7 and 1.7 a.u., and either 
short-range functional combined with long-range CCSD(T).
In the figure we see as well the reduction of the influence of the 
extrapolation procedure when going from large $\mu$ to $\mu=0$. 

For simplicity we will use for further comparisons $\mu=1.0$ a.u. at the 
lower end of the interval of confidence, however
significantly larger than the commonly employed value of 0.5 a.u. (see, e.g., 
Refs.~\cite{GerAng-CPL-05a,Goll2,Mussard}). For this value of $\mu=1.0$ a.u., 
the contribution from the short-range functional with its advantages (low 
computational cost, small basis-set superposition error, and weak 
basis-set dependence) is still present. Without loosing too much in accuracy we 
still may carry out the calculations in the aug-cc-pVTZ basis set.

\begin{figure}[h!]
  \includegraphics[width=\textwidth]{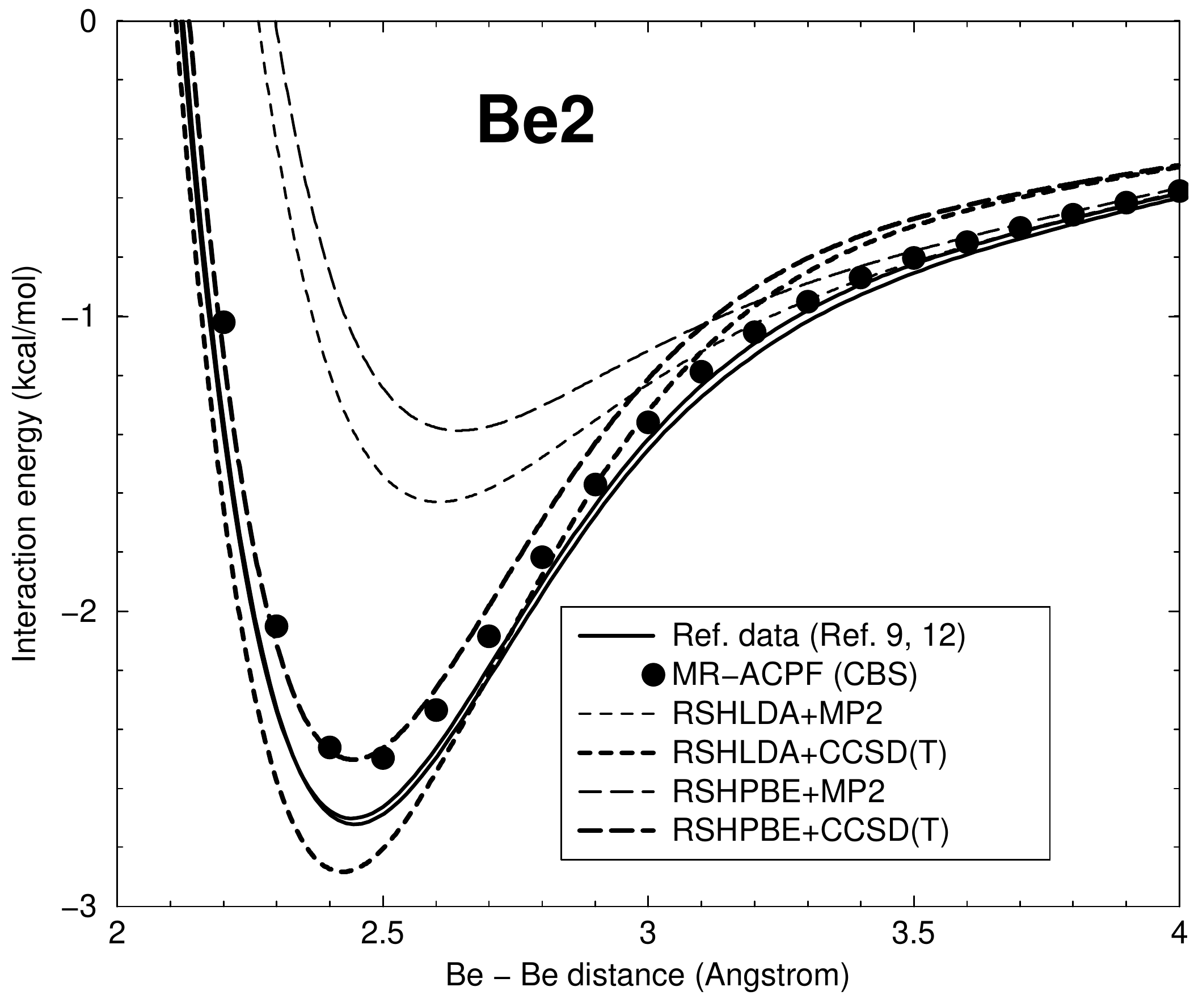}
  \caption{Interaction energy curves of the RS-DFT methods for
    the Be dimer, compared to the reference data. Basis set is the aug-cc-pVTZ
    one, and the range-separation parameter $\mu$ is taken equal to 1 a.u.. 
    Of the
    corresponding curves the RSHPBE ones lie always above the RSHLDA ones.}
  \label{fig:RSDFTpotBe2_bw}  
\end{figure}

\begin{figure}[h!]
  \includegraphics[width=\textwidth]{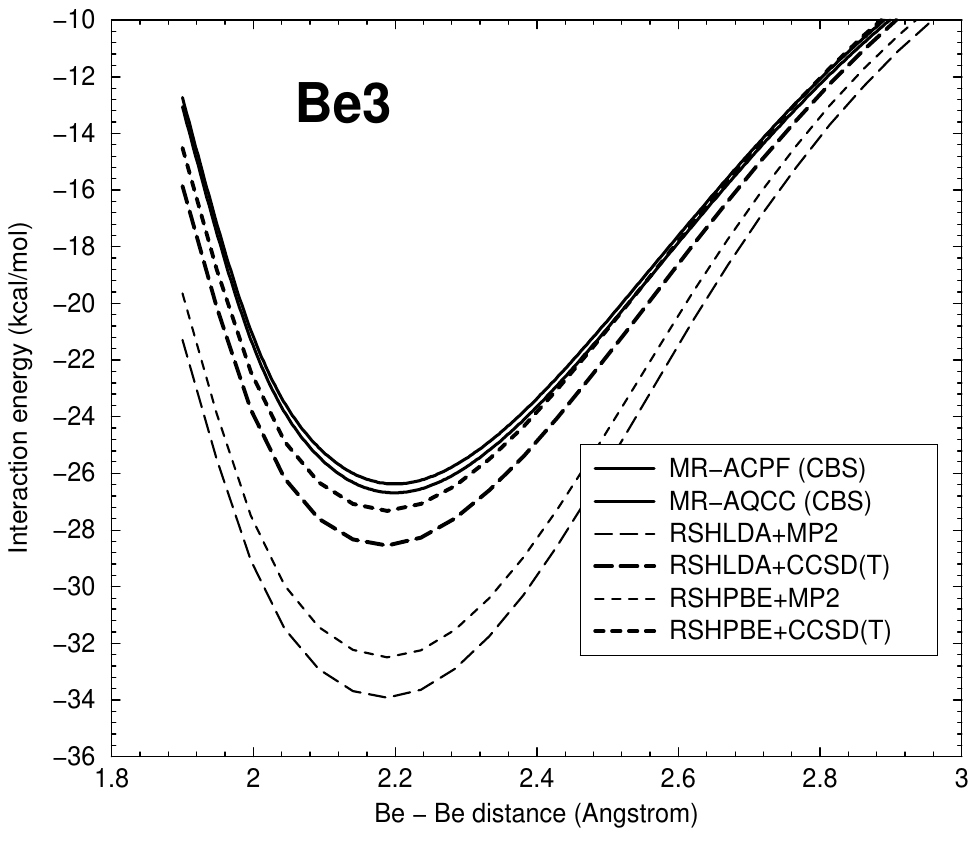}
\caption{Interaction energy curves of the RS-DFT methods for
  the Be trimer, compared to our own reference data. Same parameters as for 
  Fig.\ \ref{fig:RSDFTpotBe2_bw}. RSHLDA interaction energies are always
  smaller than the  
  RSHPBE ones.} 
\label{fig:RSDFTpotBe3_bw}  
\end{figure}

\begin{table*}[h!]
  \caption{Equilibrium bond lengths (\AA) and interaction energies (kcal/mol)
    for different methods and basis sets for the Be dimer and trimer.} 
  \begin{tabular}{l|rrrrrr}
    \hline\hline
    & \multicolumn{6}{c}{Be$_2$} \\
    &   \multicolumn{2}{c}{aug-cc-pVDZ}  & \multicolumn{2}{c}{aug-cc-pVTZ} &
    \multicolumn{2}{c}{aug-cc-pVQZ} \\ 
    & $r_{\rm Be-Be}$ & $E_{\rm min}$   & $r_{\rm Be-Be} $ & $E_{\rm min}$   &
    $r_{\rm Be-Be}$ & $E_{\rm min}$   \\ 
    \hline
    LDA              & 2.44 & $-$12.5 & 2.46 & $-$12.8  & 2.33 & $-$12.9 \\
    PBE              & 2.45 & $-$9.5  & 2.44 & $-$9.7   & 2.43 & $-$9.8 \\
    B3LYP            & 2.51 & $-$3.9  & 2.49 & $-$4.1   & 2.40 & $-$4.1 \\
    MP2              &  --- &   ---   & 2.83 & $-$0.9   & 2.74 & $-$1.2 \\
    CCSD(T)          &  --- &   ---   & 2.54 & $-$1.2   & 2.50 & $-$1.6 \\
    RSHLDA $\mu=1$   &  --- &   ---   &  --- & --- &  --- & ---  \\
    RSHPBE $\mu=1$   &  --- &   ---   &  --- & --- &  --- & ---  \\
RSHLDA+MP2 $\mu=1$   & 2.65 &  $-$1.0   & 2.61 &   $-$1.6 & 2.59 & $-$1.8 \\
RSHPBE+MP2 $\mu=1$   & 2.69 &  $-$0.8   & 2.64 &   $-$1.4 & 2.63 & $-$1.5 \\
RSHLDA+CCSD(T) $\mu=1$ & 2.45 &  $-$2.0 & 2.42 &   $-$2.9 & 2.41 & $-$3.1 \\
RSHPBE+CCSD(T) $\mu=1$ & 2.48 &  $-$1.6 & 2.44 &   $-$2.5 & 2.44 & $-$2.7 \\
\hline
& \multicolumn{6}{c}{Be$_3$}\\ 
    LDA               & 2.17 & $-$54.7   & 2.14 & $-$56.7 & 2.14 & $-$57.2 \\
    PBE               & 2.20 & $-$44.8   & 2.18 & $-$46.2 & 2.17 & $-$46.6 \\
    B3LYP             & 2.18 & $-$29.6   & 2.16 & $-$31.2 & 2.16 & $-$31.5 \\
    MP2               & 2.26 &  $-$23.0  & 2.22 & $-$28.1 & 2.21 & $-$29.7  \\
    CCSD(T)           & 2.27 &  $-$16.2  & 2.23 & $-$21.6 & 2.22 & $-$23.1  \\
    RSHLDA $\mu=1$    & 2.15 &   $-$9.2  & 2.11 & $-$10.8 & 2.11 & $-$11.2  \\
    RSHPBE $\mu=1$    & 2.15 &   $-$7.7  & 2.12 &  $-$9.0 & 2.12 & $-$9.4   \\
RSHLDA+MP2 $\mu=1$    & 2.21 &  $-$30.6  & 2.19 & $-$33.9 & 2.18 & $-$34.7  \\
RSHPBE+MP2 $\mu=1$    & 2.21 &  $-$29.3  & 2.19 & $-$32.5 & 2.18 & $-$33.2  \\
RSHLDA+CCSD(T) $\mu=1$ & 2.21 &  $-$24.9  & 2.18 & $-$28.5 & 2.18 & $-$29.3  \\
RSHPBE+CCSD(T) $\mu=1$ & 2.21 &  $-$23.8  & 2.19 & $-$27.3 & 2.18 & $-$28.0  \\
\hline
  \end{tabular}
  \label{tab:be2-be3}
\end{table*}

Figures~\ref{fig:RSDFTpotBe2_bw} and~\ref{fig:RSDFTpotBe3_bw} show the
RSHLDA+MP2, RSHPBE+MP2, RSHLDA+CCSD(T) and RSHPBE+CCSD(T) interaction 
energies with
aug-cc-pVTZ basis for $\mu=1\,$ a.u., for the dimer and for the trimer,
in comparison to the available reference data. The equilibrium bond lengths
and interaction energies obtained from RS-DFT calculations with a series of
basis sets are reported in Table~\ref{tab:be2-be3}, and compared to those
obtained from standard DFT, MP2, and CCSD(T) calculations.

Calculations with standard density-functional
approximations overestimate the interaction energies considerably. For
instance in the case of the dimer with a factor of 5 for LDA, and still a 
factor of 1.5 for B3LYP. RSHLDA+CCSD(T) and RSHPBE+CCSD(T) produce interaction 
energies in good agreement with the reference data. In contrast, RSHLDA+MP2 and 
RSHPBE+MP2 significantly underbind the dimer and overbind the trimer, as we 
have seen already from Figs. \ref{fig:RSDFTpotBe2_bw}\ and 
\ref{fig:RSDFTpotBe3_bw}. Note that, for Be$_3$, RSH+CCSD(T) (Fig.\ \ref{fig:RSDFTpotBe3_bw}) 
and CCSD(T) (Fig.\ \ref{fig:Pot_Be3_bw}) give the same deviation from the
reference data for large interatomic distances.

For the Be$_2$ case the energy differences due to different basis sets are 
small, and we may look first at the Be$_3$ results. We see that if we take 
the difference in binding energies between the DZ and TZ as one, the 
difference between TZ and QZ is about one quarter, consistently for all 
methods. In absolute values the differences due to the basis sets are divided 
by a factor of two between MP2/CCSD(T) and the RS-DFT interaction energies. 
However, using RS-DFT this dependence is twice as large as for the pure
density-functional calculations, employing the standard functionals LDA, PBE,
B3LYP or the short-range functionals only. The same factor of 4 between
the DZ--TZ and the TZ--QZ differences is found in the case of the Be$_2$ in
the RS-DFT calculations.

\begin{figure}[h!]
  \includegraphics[width=\textwidth]{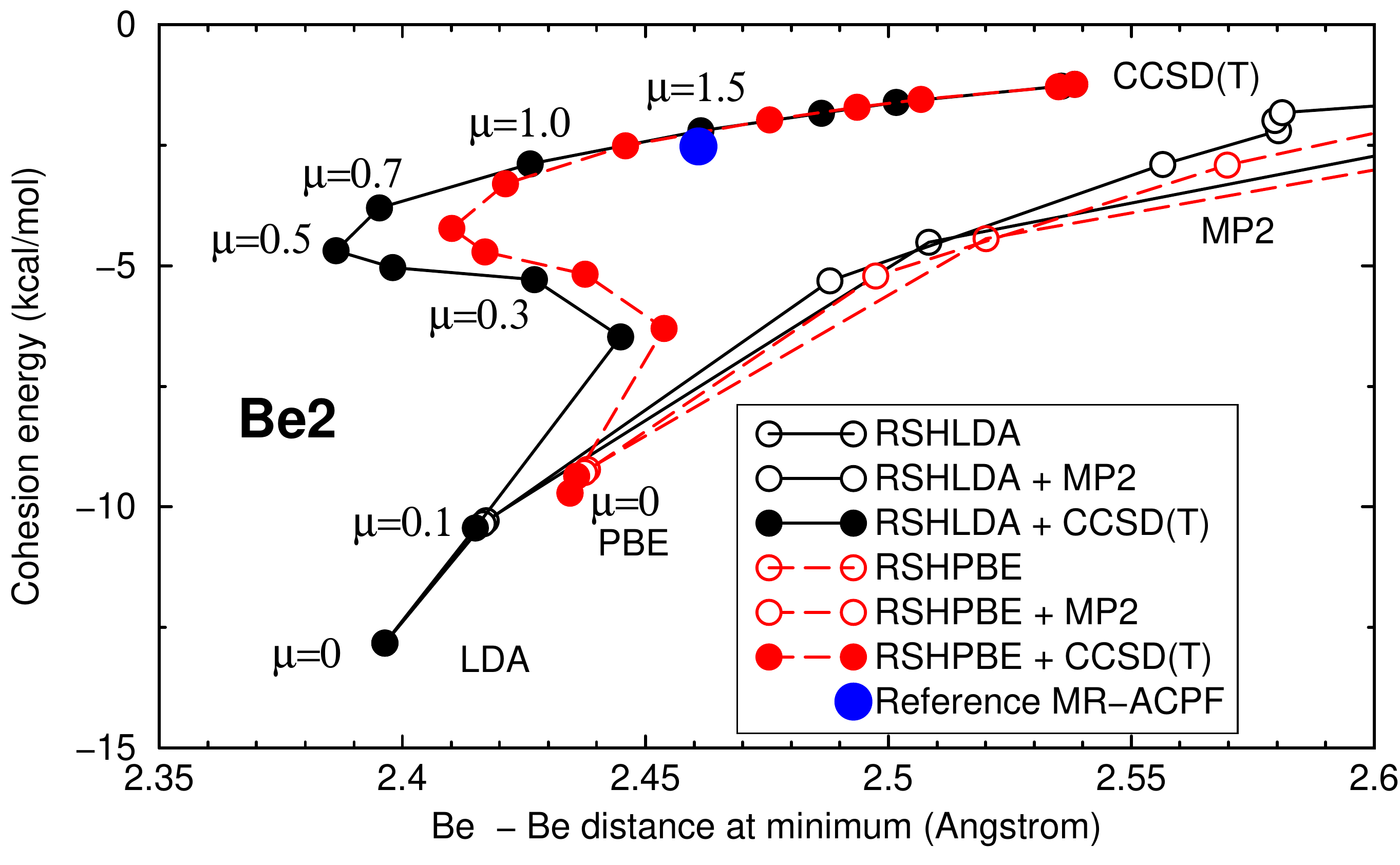}
\caption{Equilibrium cohesion energies and bond lengths as a function of the 
range-separation parameter $\mu$, for Be$_2$ and the aug-cc-pVTZ basis set. We 
include our basis-set extrapolated MR-ACPF/AQCC reference point.}
\label{fig:Be2_mu}  
\end{figure}

\begin{figure}[h!]
  \includegraphics[width=\textwidth]{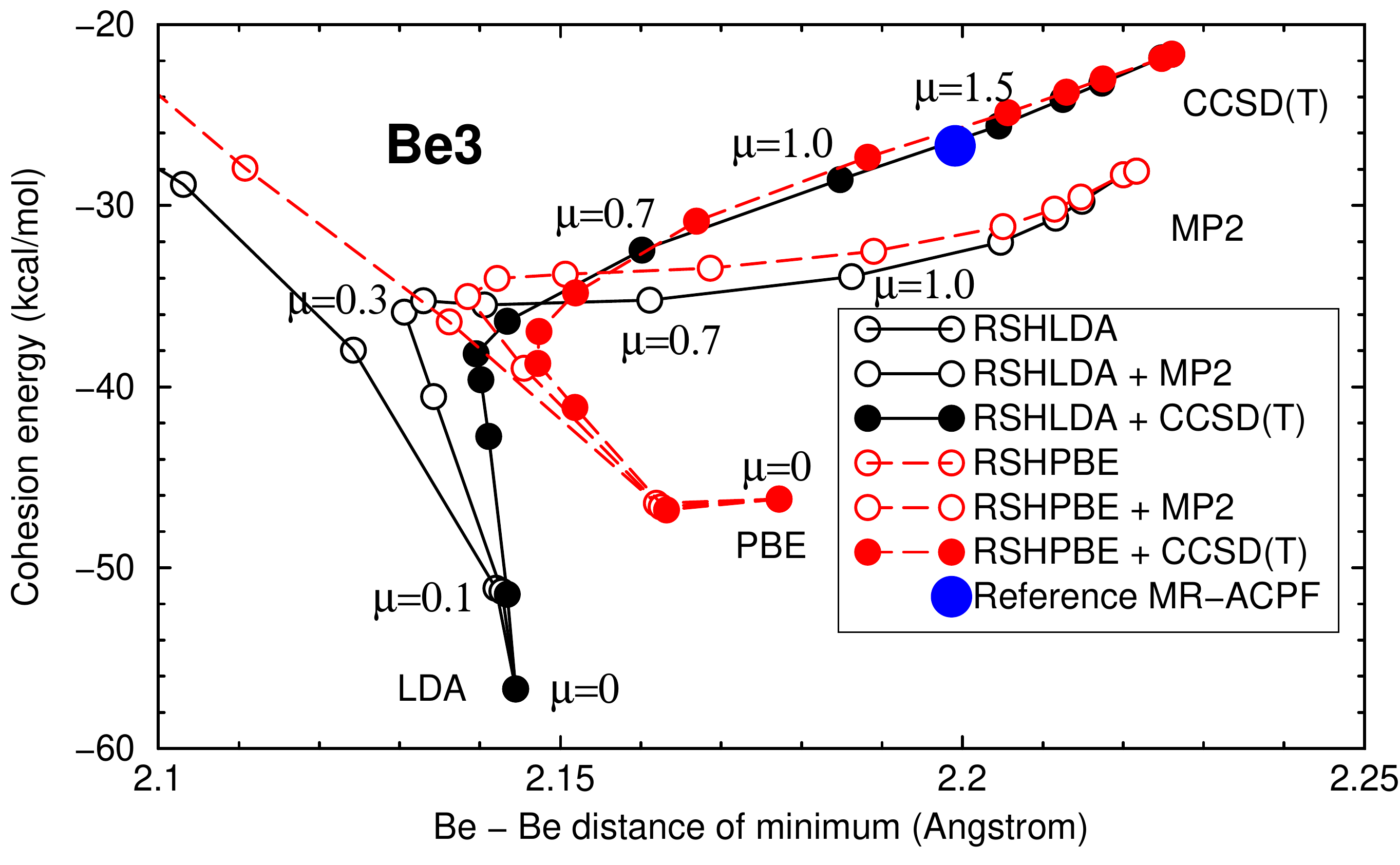}
\caption{Equilibrium cohesion energies and bond lengths as a function of the 
range-separation parameter $\mu$, for Be$_3$ and the aug-cc-pVTZ basis set.}
\label{fig:Be3_mu}  
\end{figure}

Let us look in more detail into the RS-DFT calculations (Figures 
\ref{fig:Be2_mu} and \ref{fig:Be3_mu}). Varying the range-separation parameter, 
we determine the minima of the cohesion-energy curves, which all lie in a 
narrow range of distances, but not on energies. Again we note that MP2 and 
CCSD(T) long-range correlations have not the same effect on the results. Indeed,
when aiming at reproducing the reference calculations, electron correlation of 
order higher than second-order perturbation theory is needed. This is more 
pronounced in Be$_2$ than in Be$_3$, where RSH+MP2 or RSH+CCSD(T) follow 
similar trends with the range-separation parameter $\mu$. The interpolation 
between $\mu=0$ (pure Kohn-Sham DFT) and $\mu\rightarrow\infty$ (no 
contribution of a functional) appears to be not at all linear, neither for 
Be$_2$ nor Be$_3$, and the optimal range-separation parameter should 
be chosen significantly larger than 0.5 a.u., as found before. Of course, for 
large  $\mu$ the difference between the two functionals employed becomes 
reasonably small. 

Before concluding we might look at the performances of other recent DFT 
approaches as provided by the double-hybrid functionals including both 
HF exchange and a MP2 correlation part. Thus results should lie 
somewhere between standard MP2 and our RSH+MP2. Using the 
B2PLYP \cite{B2PLYP}, the XYG3 \cite{XYG3}, and the $\omega$B97X-2(TQZ) 
\cite{wB97X2} functionals we find indeed (data in the Supplementary Material) 
for the trimer that all functionals overshoot as did MP2 and RSH+MP2 already. 
For the dimer the situation is a little different as MP2 and RSH+MP2 resulted 
in too weak an interaction, but the double hybrids overbind, and the three
variants lead to significantly different shapes of the potential.
 
\section{Conclusion}

We investigated the challenging systems Be$_2$ and Be$_3$ with different
wave-function and DFT methods, comparing these to 
RS-DFT approaches, which are capable to describe explicitly
long-range correlations, absent in standard Kohn-Sham theory with semilocal 
approximations. The interaction energy in small Be clusters is certainly not 
only due to dispersion as the interatomic distances are far smaller than typical
van-der-Waals interactions, and interaction energies are much higher.

We observe that for this particular type of bonding -- no chemical bond
properly speaking, but no dispersion-only binding either (as for rare-gas
complexes), the RS-DFT approach, when used with the usual range-separation 
parameter of $\mu=0.5\,$a.u., produces too high binding energies, similar to 
the commonly employed B3LYP functional. Using a value of $\mu$ around 1 a.u. in 
our RS-DFT scheme with a CCSD(T) long-range part permits to reproduce well the 
MR--ACPF or MR--AQCC reference energies of the trimer.

Long-range single-reference CCSD(T) thus seems adequate 
for obtaining a reliable binding energy -- even if perhaps due to fortuitous 
error cancellation. On the other hand, adding only second-order diagrams for 
expanding the long-range correlation energy is not sufficient, for any value 
of the range-separation parameter. This last point is in particular 
important for the modern development of double hybrids, leaving room for 
further developments.

\section{Acknowledgments}
This work has been partly financed by the french ANR (Agence Nationale de la
Recherche) through contract Nr.\ ANR 07-BLAN-0272 (project Wademecom). The
authors thank D.\ Andrae (Berlin), H.\ Stoll (Stuttgart) and F.\ Spiegelmann 
(Toulouse) for fruitful hints. Computer time has been provided by the 
F\'ed\'eration de Recherche IP2CT (Sorbonne University, Paris).

\appendix

\section{Technical details}
All calculations were performed on 3.3 GHz Intel PCs running
CentOS Linux, using Molpro \cite{Molpro} in its version 2008.2 (standard DFT
and MCSCF/MRCI calculations) and a local development version, 2008.2, 
for the RS-DFT calculations. The employed aug-cc-pVXZ basis sets originate 
from the Gaussian basis set exchange form at Pacific National Laboratories.
Only for the calculations on double hybrids we employed the QChem code, version 
5.0 \cite{QChem}.

For the single- and multi-reference methods and the long-range correlation
part of the RS-DFT calculations only the valence electrons of the Be atom were
correlated. Energy thresholds were 10$^{-7}\,$a.u.\ for convergence and
10$^{-9}\,$a.u.\ as target accuracy for establishing the grid for the
numerical density-functional integration.   

The multi-reference calculations started from a complete-active-space (CAS) 
wave function with the two valence electrons in four orbitals for each
atom. The subsequent ACPF or AQCC calculations used this CAS wave function as
reference space (60 configuration state functions (CSFs) for Be$_2$, and about
4000 CSFs for Be$_3$). 

Minima in the 1D potential curves were determined through a spline fit with
points spaced at 0.1$\,$\AA, starting at 1.9$\,$\AA\ and including some large
distances to check size consistency.


\begin{thebibliography}{0}%
\makeatletter
\providecommand \@ifxundefined [1]{%
 \@ifx{#1\undefined}
}%
\providecommand \@ifnum [1]{%
 \ifnum #1\expandafter \@firstoftwo
 \else \expandafter \@secondoftwo
 \fi
}%
\providecommand \@ifx [1]{%
 \ifx #1\expandafter \@firstoftwo
 \else \expandafter \@secondoftwo
 \fi
}%
\providecommand \natexlab [1]{#1}%
\providecommand \enquote  [1]{``#1''}%
\providecommand \bibnamefont  [1]{#1}%
\providecommand \bibfnamefont [1]{#1}%
\providecommand \citenamefont [1]{#1}%
\providecommand \href@noop [0]{\@secondoftwo}%
\providecommand \href [0]{\begingroup \@sanitize@url \@href}%
\providecommand \@href[1]{\@@startlink{#1}\@@href}%
\providecommand \@@href[1]{\endgroup#1\@@endlink}%
\providecommand \@sanitize@url [0]{\catcode `\\12\catcode `\$12\catcode
  `\&12\catcode `\#12\catcode `\^12\catcode `\_12\catcode `\%12\relax}%
\providecommand \@@startlink[1]{}%
\providecommand \@@endlink[0]{}%
\providecommand \url  [0]{\begingroup\@sanitize@url \@url }%
\providecommand \@url [1]{\endgroup\@href {#1}{\urlprefix }}%
\providecommand \urlprefix  [0]{URL }%
\providecommand \Eprint [0]{\href }%
\providecommand \doibase [0]{http://dx.doi.org/}%
\providecommand \selectlanguage [0]{\@gobble}%
\providecommand \bibinfo  [0]{\@secondoftwo}%
\providecommand \bibfield  [0]{\@secondoftwo}%
\providecommand \translation [1]{[#1]}%
\providecommand \BibitemOpen [0]{}%
\providecommand \bibitemStop [0]{}%
\providecommand \bibitemNoStop [0]{.\EOS\space}%
\providecommand \EOS [0]{\spacefactor3000\relax}%
\providecommand \BibitemShut  [1]{\csname bibitem#1\endcsname}%
\let\auto@bib@innerbib\@empty
\end{thebibliography}%


\begin{thebibliography}{}

\bibitem{rangeseparate}\reference{\'Angy\'an JG, Gerber IC, Savin A,
        Toulouse J}{Phys.Rev.}{A 72}{2005}{012510}

\bibitem{ToulouseRPA}\reference{Toulouse J,Zhu W, Angyan JG, Savin A}{Phys.Rev.}{A82}{2010}{032502}

\bibitem{ToulouseCCD}\reference{Toulouse J, Zhu W, Savin A, Jansen G, \'Angy\'an JG}{J.Chem.Phys.}{135}{2011}{084119}

\bibitem{Chermak}\reference{Chermak E, Mussard B, \'Angy\'an JG, Reinhardt P}{Chem.Phys.Lett.}{550}{2012}{162}

\bibitem{Mussard}\reference{Mussard B, Reinhardt P, \'Angy\'an JG, Toulouse J}{J.Chem.Phys.}{142}{2015}{154123}

\bibitem{JonesBe2}\reference{Jones RO}{J.Chem.Phys.}{71}{1979}{1300}


\bibitem{LiuMcLean}\reference{Liu B, McLean AD} {J.Chem.Phys.}
  {72}{1980}{3418} 

\bibitem{BeExp}\reference{Bondybey VE}{Chem.Phys.Lett.}{109}{1984}{436} 

\bibitem{Roeggen}\reference{R\o{}eggen I, Veseth
    L}{Int.J.Quant.Chem.}{101}{2005}{201}  

\bibitem{PatkowskiBeMgCa}\reference{Patkowski K, Podeszwa R,
    Szalewicz K}{J.Phys.Chem.A}{111}{2007}{12822}   

\bibitem{Ruedenberg}\reference{Schmidt MW, Ivanic J, Ruedenberg
    K}{J.Phys.Chem.A}{114}{2010}{8687}

\bibitem{Patkowski}\reference{Patkowski K, Spirko V,
    Szalewicz K}{Science}{326}{2009}{1382}   

\bibitem{ExpBeII}\reference{Merritt JM, Bondybey VE, Heaven
    MC}{Science}{324}{2009}{1548}

\bibitem{Kalemos}\reference{Kalemos A}{J.Chem.Phys.}{145}{2016}{214302}


\bibitem{KohnSham}\reference{Kohn W, Sham LJ}{Phys.Rev.}{140}{1965}{A1133}   

\bibitem{Colonna}\reference{Toulouse J, Colonna F,
    Savin A}{Phys.Rev.A}{70}{2004}{062505}

\bibitem{Grimme}\reference{Grimme S}{J.Comp.Chem.}{25}{2004}{1463}

\bibitem{Elstner}\reference{Elstner M, Hobza P, Frauenheim T, Suhai S,
  Kaxiras E}{J.Chem.Phys.}{114}{2001}{5149} 

\bibitem{Johnson}\reference{Becke AD,
    Johnson ER}{J.Chem.Phys.}{122}{2005}{154104} 

\bibitem{MollerPlesset}\reference{M\o{}ller C,
    Plesset MS}{Phys.Rev.}{46}{1934}{618}  

\bibitem{Goll}\reference{Goll E, Werner HJ, Stoll H}{Phys.Chem.Chem.Phys.}
     {7}{2005}{3917}

\bibitem{Goll2}\reference{Goll E, Werner HJ, Stoll H, Leininger T, Gori-Giorgi P, Savin A}{Chem.Phys.}
     {329}{2006}{276}

\bibitem{SRfunctionals}\reference{Toulouse J, Savin A, Flad
    HJ}{Intern.J.Quant.Chem.}{100}{2004}{1047} 

\bibitem{ColonnaII}\reference{Toulouse J, Colonna F, Savin
    A}{J.Chem.Phys.}{122}{2005}{014110} 

\bibitem{Paziani}\reference{Paziani S, Moroni S, Gori-Giorgi P, 
    Bachelet GB}{Phys.Rev.B}{73}{2006}{155111}   

\bibitem{muhalfFro}\reference{Fromager E, Toulouse J,
    Jensen HJ Aa}{J.Chem.Phys.}{126}{2007}{074111} 

\bibitem{GollTPSS}\reference{Goll E, Ernst M, Moegle-Hofacker F,
    Stoll H}{J.Chem.Phys.}{130}{2009}{234112}

\bibitem{BeReview}\reference{Heaven MC, Merritt JM, Bondybey 
VE}{Ann.Rev.Phys.Chem.}{62}{2011}{375}

\bibitem{HandyI}\reference{Harrison RJ,
    Handy NC}{Chem.Phys.Lett.}{98}{1983}{97}  

\bibitem{Lepetit}\reference{Lepetit MB, Malrieu
    JP}{Chem.Phys.Lett.}{169}{1990}{285} 

\bibitem{Zhang}\reference{Zhang H, Ma J, Reinhardt P, 
    Malrieu JP}{J.Chem.Phys}{132}{2009}{034108} 

\bibitem{SCSC}\reference{Daudey JP, Heully JL,
    Malrieu JP}{J.Chem.Phys.}{99}{1993}{1240}

\bibitem{Bartlett}\reference{Sosa C, Noga J,
    Bartlett RJ}{J.Chem.Phys.}{88}{1988}{5974} 

 \bibitem{BeAngyan}\reference{Gerber IC,
     \'Angy\'an JG}{Chem.Phys.Lett.}{416}{2005}{370} 

 \bibitem{ToulousePRL}\reference{Toulouse J, Gerber IC, Jansen G, Savin A,
     \'Angy\'an JG}{Phys.Rev.Lett.}{102}{2009}{096404} 

\bibitem{Fromager}\reference{Fromager E, Chimiraglia R,
    Jensen HJ Aa}{Phys.Rev.A}{81}{2010}{024502}

\bibitem{HandyII}\reference{Harrison RJ,
    Handy NC}{Chem.Phys.Lett.}{123}{1986}{321}  

\bibitem{Taylor}\reference{Lee TJ, Rendell AP,
    Taylor PR}{J.Chem.Phys.}{92}{1990}{489} 

\bibitem{Kaplan}\reference{Kaplan IG, Roszak S,
    Leszczynski J}{J.Chem.Phys.}{113}{2000}{6245} 

\bibitem{Klopper}\reference{Klopper W,
    Alml\"of J}{J.Chem.Phys.}{99}{1993}{5167} 

\bibitem{Stefano}\reference{Junquera-Hern\'andez JM, S\'anchez-Mar\'{\i}n J, 
    Bendazzoli GL, Evangelisti S}{J.Chem.Phys.}{120}{2004}{8405} 

\bibitem{Novaro}\reference{Novaro O, Ko\l{}os W}{J.Chem.Phys.}{67}{1967}{5066}  

\bibitem{DaudeyBe}\reference{Daudey JP, Novaro O, Ko\l{}os W,
    Berrondo M}{J.Chem.Phys.}{71}{1979}{4297} 

\bibitem{Jae}\reference{Jae SL}{Phys.Rev.A}{68}{2003}{43201}

\bibitem{Khanna}\reference{Khanna SN, Reuse F,
    Buttet J}{Phys.Rev.Lett.}{61}{1988}{535} 

\bibitem{LDABe3}\reference{Rao BK, Khanna SN, Meng J,
    Jena P}{Z.Phys.D}{18}{1991}{171} 

\bibitem{Jellinek}\reference{Sriniva S, Jellinek J}{J.Chem.Phys.}{121}
  {2004}{7243} 

\bibitem{Beyer}\reference{Beyer MK, Kaledin LA, Kaledin AL, Heaven MC, 
    Bondybey VE}{Chem.Phys.}{262}{2000}{15} 

\bibitem{Gdanitz}\reference{Gdanitz R, Ahlrichs R}{Chem.Phys.Lett.}{143}{1988}{413}
  
\bibitem{AQCC}\reference{Meissner L}{Chem.Phys.Lett.}{146}{1988}{205};
  \reference{Szalay PG, Bartlett RL}{Chem.Phys.Lett.}{214}{1993}{481};
  \reference{F\"usti-Moln\'ar L, Szalay PG}{Chem.Phys.Lett.}{258}{1996}{400} 

\bibitem{Molpro} Molpro 2006.1 and 2008.2; {\tt MOLPRO} is a package of
  ab-initio 
  programs written by {\sl Werner HJ and Knowles PJ} with contributions
  from {\sl Alml\"of J, Amos RD, Bernhardsson A, Berning A, Cooper DL, 
    Deegan MJO, Dobbyn AJ, Eckert F, Hampel C, Lindh R, Lloyd AW, 
    Meyer W, Mura ME,  Nicklass A, Peterson K, Pitzer R, Pulay P,
  Rauhut G, Sch\"utz M, Stoll H, Stone AJ, Taylor PR and Thorsteinsson T}, 
  University of Stuttgart and Birmingham, 1998---

\bibitem{Odile}\reference{Franck O, Mussard B, Luppi E, Toulouse 
J}{J.Chem.Phys.}{142}{2015}{074107}

\bibitem{BoysBernardi}\reference{Boys SF,
    Bernardi F}{Mol.Phys.}{19}{553}{1970}

\bibitem{CPP}\reference{Stoll H, Fuentealba P, Schwerdtfeger P, Flad J,
        v.\ Szentp\'aly L, Preuss HW}{J.Chem.Phys.}{81}{1984}{2732}

\bibitem{Peterson}{\sl Peterson K} private communication (2009)

 \bibitem{GerAng-CPL-05a}\reference{Gerber IC,
     \'Angy\'an JG}{Chem.Phys.Lett.}{415}{2005}{100} 













\bibitem{B2PLYP}\reference{Grimme S}{J. Chem. Phys.}{124}{2009}{034108}
\bibitem{XYG3}\reference{Zhang Y, Xu X, Goddard III WA}{Proc. Natl. Acad. Sci. USA}{106}{2009}{4963}
\bibitem{wB97X2}\reference{Chai J-D, Head-Gordon M}{J. Chem. Phys.}{131}{2009}{174105}

\bibitem{QChem}\reference{Shao Y, Gan Z, Epifanovsky E, Gilbert ATB, Wormit M, 
Kussmann J, Lange AW, Behn A, Deng J, Feng X, Ghosh D, Goldey M, Horn PR, 
Jacobson LD, Kaliman I, Khaliullin RZ, K\'us T, Landau A, Liu J, Proynov EI, 
Rhee YM, Richard RM, Rohrdanz MA, Steele RP, Sundstrom EJ, Woodcock III HL, 
Zimmerman PM, Zuev D, Albrecht B, Alguire E, Austin B, Beran GJO, Bernard YA, 
Berquist E, Brandhorst K, Bravaya KB,  Brown ST, Casanova D, Chang CM, Chen Y, 
Chien SH, Closser KD, Crittenden DL, Diedenhofen M, DiStasio Jr. RA, Dop H, 
Dutoi AD, Edgar RG, Fatehi S, F\"usti-Molnar L, Ghysels A, Golubeva-Zadorozhnaya 
A, Gomes J, Hanson-Heine MWD, Harbach PHP, Hauser AW, Hohenstein EG, Holden ZC, 
Jagau TC, Ji H, Kaduk B, Khistyaev K, Kim J, King RA, Klunzinger P,Kosenkov D, 
Kowalczyk T, Krauter CM, Lao KU, Laurent A, Lawler KV, Levchenko SV, Lin CY, Liu 
F, Livshits E, Lochan RC, Luenser A, Manohar P, Manzer SF, Mao SP, Mardirossian 
N, Marenich AV, Maurer SA, Mayhall NJ, Oana CM, Olivares-Amaya R, O’Neill DP, 
Parkhill JA, Perrine TM, Peverati R, Pieniazek PA, Prociuk A, Rehn DR, Rosta E, 
Russ NJ, Sergueev N, Sharada SM, Sharmaa S, Small DW, Sodt A, Stein T, St\"uck 
D, Su YC, Thom AJW, Tsuchimochi T, Vogt L, Vydrov O, Wang T, Watson MA, Wenzel 
J, White A, Williams CF, Vanovschi V, Yeganeh S, Yost SR, You ZQ, Zhang IY, 
Zhang X, Zhou Y, Brooks BR, Chan GKL, Chipman DM, Cramer CJ, Goddard III WA, 
Gordon MS, Hehre WJ, Klamt A, Schaefer III HF, Schmidt MW, Sherrill CD, Truhlar 
DG, Warshel A, Xua X, Aspuru-Guzik A, Baer R, Bell AT, Besley NA, Chai JD, Dreuw 
A, Dunietz BD, Furlani TR, Gwaltney SR, Hsu CP, Jung Y, Kong J, Lambrecht DS, 
Liang W, Ochsenfeld C, Rassolov VA, Slipchenko LV, Subotnik JE, Van Voorhis T, 
Herbert JM, Krylov AI, Gill PMW, and Head-Gordon M}{Mol. Phys.}{113}{2015}{184}



\end{thebibliography}
\end{document}